\documentclass[prb,showpacs,twocolumn,amsmath,amssymb]{revtex4}% Physical Review B

\usepackage{graphicx}% Include figure files
\usepackage{dcolumn}% Align table columns on decimal point
\usepackage{bm}% bold math

\newcommand{\Vvec}[1]{\mbox{\boldmath$#1$}}
\newcommand{\Vvecs}[1]{\mbox{\scriptsize\boldmath$#1$}}

\begin{document}

%\preprint{APS/123-QED}

\title{Transport phenomena in three-dimensional system close to the
magnetic quantum critical point: The conserving approximation with the current vertex corrections}

\author{Seiichiro Onari}
%\altaffiliation[Present address: ]{Department of Applied Physics, Nagoya University, Chikusa, Nagoya 464-8603, Japan.}
\author{Hiroshi Kontani$^1$}
%\altaffiliation[On leave of absence from ]{Department of Physics, University of Tokyo.}
\author{Yukio Tanaka}
\affiliation{Department of Applied Physics, Nagoya University, Chikusa, Nagoya 464-8603, Japan.}
\affiliation{$^1$Department of Physics, Nagoya University, Chikusa, Nagoya 464-8603, Japan.}

\date{\today}% It is always \today, today,
             %  but any date may be explicitly specified

\begin{abstract}
It is known that various transport coefficients strongly
 deviate from conventional Fermi-liquid behaviors
 in many electron systems which are
 close to antiferromagnetic (AF) quantum critical points (QCP). For example, Hall coefficients and Nernst coefficients in three-dimensional heavy fermion CeCoIn$_5$ and
 CeCu$_{6-x}$Au$_x$ increase strikingly at low temperatures, 
whose overall behaviors are similar to those in high-$T_c$
 cuprates. These temperature dependences are too strong to explain in
 terms of the relaxation time approximation.
To elucidate the origin of these anomalous transport phenomena in
 three-dimensional systems, we study the current vertex
 corrections (CVC) based on the fluctuation exchange (FLEX)
 approximation, and find out decisive role of the CVC. 
The main finding of the present paper is that the Hall
 coefficient and the Nernst coefficient strongly increase thanks to the
 CVC in the vicinity of the AF QCP, irrespective of dimensionality.
We also study the relaxation time of quasi-particles, and find that
 ``hot points'' and ``cold lines'' are formed in general
 three-dimensional systems
due to strong AF fluctuations.
\end{abstract}


%Use showkeys class option if keyword
                              %display desired
\maketitle

\section{INTRODUCTION}
Strongly correlated electron systems close to the quantum critical point (QCP) have stimulated
much interest.
Especially, heavy fermion compound CeCoIn$_5$,\cite{petrovic} which is a
three-dimensional metal close to the antiferromagnetic (AF) QCP,
attracts much attention because of the non-Fermi-liquid normal state and the
$d$-wave superconductivity ($T_c=2.3$K).
Moreover, recent experimental efforts have revealed the existence of interesting
anomalous transport phenomena characteristic of the AF QCP. In the normal
state of CeCoIn$_5$, for example, it is observed that the resistivity\cite{nakajima} $\rho
\propto T$, the Hall coefficient\cite{nakajima} $R_H\propto 1/T$, and the
Nernst coefficient\cite{bel} $\nu\propto 1/T$ below $20$K till
$T_c=2.3$K. These behaviors are quite different from the normal
Fermi-liquid behaviors, $\rho\propto T^2$ and $R_H\propto T^0$.

In Ce115, the maximum values of $R_H$ and $\nu$ are quite huge compared to
values for high temperatures. The maximum value of $R_H$ becomes about
$30$ times larger than the values at high temperatures. 
The value of $\nu$ reaches
$-1.0\mu {\rm V/KT}$ at $5$K, whose magnitude is about $1000$ times larger than
the values in usual metals. The temperature dependence of $R_H$ is similar to the two-dimensional high-$T_c$
cuprates above the pseudo-gap temperatures, and that of $\nu$ is similar
to electron doped high-$T_c$ cuprates, irrespective of signs. Moreover,
the magnitude of $R_H$ and $\nu$ in CeCoIn$_5$ is much larger than that
in high-$T_c$ cuprates. Similar drastic increases of $R_H$ are also
observed in CeCu$_{6-x}$Au$_x$\cite{fukuhara} and
YbRh$_2$Si$_2$\cite{paschen}, which are three-dimensional heavy fermion
compounds close to the AF QCP.

The relaxation time
approximation (RTA)\cite{stojkovic,rosch} has been used
frequently in the study of transport phenomena, although its
reliability for strongly correlated systems is not assured.
According to the spin fluctuation theory, 
the relaxation time
$\tau_{\Vvecs{k}}$ becomes strongly anisotropic.\cite{hlubina-rice,stojkovic} The spots on the Fermi
surface where 
$\tau_{\Vvecs{k}}$ takes the maximum (minimum) value is denoted by the cold (hot)
spots in literatures. The ratio of the relaxation time at cold spots and hot spots
$r=\tau_{\rm cold}/\tau_{\rm hot}$ and the weight of the cold spots play
an important role in the transport phenomena.
However, in terms of the RTA, an unrealistic huge $r$ (say $r=100$-$1000$) is
required to reproduce the experimental enhancement of $R_H$ in
CeCoIn$_5.$\cite{nakajima}
If we assume that $R_H$ is enhanced by this mechanism, $R_H$ should be
suppressed quite sensitively by a very small amount of impurity.
In addition, when $r\gg 1$, the magnetresistance should be too large to
explain the modified Kohler's rule, $\Delta \rho/\rho \propto
R_H^2/\rho^2$, which is observed in high-$T_c$ cuprates\cite{ong3,ando}
and in CeCoIn$_5$\cite{nakajima}.

These anomalous transport phenomena close to the AF QCP are well reproduced
by taking the current vertex corrections (CVC) into account.
Actually, the CVC is necessary to satisfy conservation laws. In the Fermi liquid theory, the CVC corresponds to the backflow, which naturally arises from electron-electron correlations. Then, the CVC is indispensable to calculate the transport coefficients in the strongly correlated electron systems, where electron-electron correlations are dominant.
For example, the modified Kohler's rule is explained due to the CVC caused
by the AF fluctuations.\cite{kontani-mag}
The negative $R_H$ in electron doped high-$T_c$
cuprates, which cannot be explained by the RTA because the Fermi surface
is hole-like everywhere, is explained if we take the CVC. 
Moreover, it is not easy to explain the enhancement of $\nu$ by the RTA
because the Sondheimer cancellation\cite{sondheimer} makes $\nu$ small.
The enhancement of $\nu$ in electron-doped (hole-doped) high-$T_c$
cuprates are caused by the CVC due to the AF (AF and superconducting) fluctuations.

Until now, various non-Fermi-liquid behaviors of high-$T_c$
cuprates have been explained by the spin fluctuation model, such as the
self-consistent renormalization (SCR) theory\cite{moriya-takahashi,ueda-moriya,moriya-ueda} and the fluctuation
exchange (FLEX) theory\cite{bickers1,bickers2,monthoux}.
For example, an appropriate behavior of the spin susceptibility and the AF correlation length $\xi$
($\xi^2\propto 1/T$)
are obtained. Moreover, spin-fluctuation theories\cite{stojkovic,moriya-ueda}
derive the relation $\rho\propto T^2\xi^2$, which is consistent with the
non-Fermi-liquid behaviors of high-$T_c$ cuprates close to the AF QCP.
Based on the spin fluctuation theory, Kontani {\it et al.}\cite{kontani-Hall,kontani-mag,kontani-ther,kontani-nu} have
developed a theory of transport phenomena, by focusing on
crucial role of the CVC.
This framework naturally reproduces the temperature dependence of
transport coefficients for high-$T_c$ cuprates and other 2D systems\cite{kontani-kino}
close to the QCP. 

In high-$T_c$ cuprates\cite{kontani-Hall}, the CVC plays an important role
on transport phenomena.
However, it is highly non-trivial whether the CVC
is significant in three-dimensional systems, since the CVC totally
vanishes in the dynamical mean field theory (DMFT) where $d=\infty$
limit is taken.\cite{kotliar,kotliar-vollhardt}
The reason is that the irreducible four-point vertex $\Gamma^{\rm I}$ in
the DMFT becomes a
local function, which cannot contribute to the CVC.
Although it is generally believed that the DMFT works well in various 3D
systems with strong correlation,\cite{kotliar} the momentum dependences
of the self-energy and the vertex corrections are significant near the
QCP. However, we must consider the CVC which cannot be taken into account within the DMFT.
To elucidate this issue, we study the role of the CVC based on the AF
fluctuation theory in three-dimensional systems.

The purpose of this paper is to examine whether non-Fermi-liquid
behaviors in 3D systems can be explained by taking account of the CVC in
terms of the FLEX approximation. In 3D systems, massive calculation
resource and time are needed to perform the calculation.
We show that striking increase of $R_H$ and $\nu$ can be obtained at low
temperatures even in 3D by virtue of the CVC, which is consistent with experiments in CeCoIn$_5$ and
CeCu$_{6-x}$Au$_x$. The present study is the first
microscopic calculation for the Hall coefficient and the Nernst
coefficient in 3D with the CVC. We
also find that in 3D systems, the hot and cold spots form
point-like (``hot points'') and line-like shape (``cold lines''), respectively.
The CVC on the cold lines plays a major role for increasing $R_H$ and $\nu$.

\section{Formulation}
\subsection{Model}
\begin{figure}[htdp]
\begin{center}
\includegraphics[height=50mm]{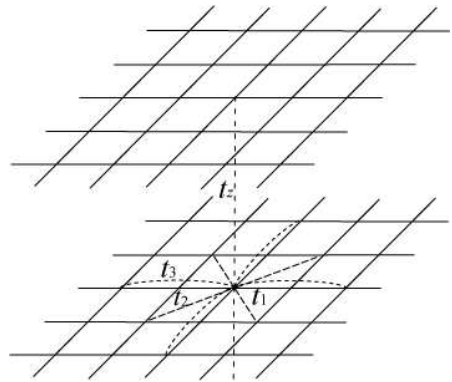}
\caption{The lattice structure with the nearest intralayer hopping
 $t_1$, the second one $t_2$, the third one $t_3$ and the interlayer
 hopping $t_z$}
\label{lattice}
\end{center}
\end{figure}

We first introduce the three-dimensional Hubbard model,
\begin{equation}{\cal H}=-\sum_{i,j}\sum_{\sigma}t_{ij}c_{i\sigma}^{\dagger}c_{j\sigma}+\frac{1}{2}\sum_{i,\sigma}Un_{i\sigma}n_{i-\sigma}
\end{equation}
on a stacked square lattice with the Coulomb repulsion $U$ and the intralayer hopping
$t_1$, $t_2$, $t_3$ and the interlayer hopping $t_z$ depicted in
Fig. \ref{lattice}.
Hereafter, we take $t_1=1$ as a unit of energy.

Then, the dispersion is derived as, 
\begin{eqnarray}
\epsilon^0_{\Vvecs{k}}=-2t_1(\cos(k_x)+\cos(k_y))-4t_2\cos(k_x)\cos(k_y)\nonumber\\
-2t_3(\cos(2k_x)+\cos(2k_y))-2t_z\cos(k_z).
\end{eqnarray}

We apply the FLEX
approximation\cite{bickers1,bickers2,koikegami,kontani} where the
Green's function, the self-energy and the susceptibility are obtained self-consistently.
The FLEX approximation belongs to ``conserving approximations''
formulated by Baym and Kadanoff\cite{baym-kadanoff,baym}.
The spin ($\chi^{\rm s}_{\Vvecs{q}}$) and charge ($\chi^{\rm c}_{\Vvecs{q}}$) susceptibilities
are
\begin{equation}
\chi^{\rm s}_{\Vvecs{q}}(\omega_l)=\frac{\chi^0_{\Vvecs{q}}(\omega_l)}{1-U\chi^0_{\Vvecs{q}}(\omega_l)},
\end{equation}
\begin{equation}
\chi^{\rm c}_{\Vvecs{q}}(\omega_l)=\frac{\chi^0_{\Vvecs{q}}(\omega_l)}{1+U\chi^0_{\Vvecs{q}}(\omega_l)},
\end{equation}
and the irreducible susceptibility $\chi^0_{\Vvecs{q}}$ is
\begin{equation}
\chi^0_{\Vvecs{q}}(\omega_l)=-\frac{T}{N}\sum_{{\Vvecs{k}},n}G_{\Vvecs{q}+\Vvecs{k}}(\omega_l+\epsilon_n)G_{\Vvecs{k}}(\epsilon_n),
\end{equation}
where Matsubara frequencies are denoted by $\epsilon_n=(2n+1)\pi T$ and
$\omega_l=2l\pi T$, respectively.

The self-energy is given by 
\begin{equation}
\Sigma_{\Vvecs{k}}(\epsilon_n)=\frac{T}{N}\sum_{{\Vvecs{q}},l}G_{\Vvecs{k}-\Vvecs{q}}(\epsilon_n-\omega_l)V_{\Vvecs{q}}(\omega_l),
\end{equation}
where the effective interaction $V_{\Vvecs{q}}$ is
\begin{equation}
V_{\Vvecs{q}}(\omega_l)=U^2\left(\frac{3}{2}\chi^{\rm s}_{\Vvecs{q}}(\omega_l)+\frac{1}{2}\chi^{\rm c}_{\Vvecs{q}}(\omega_l)-\chi^{0}_{\Vvecs{q}}(\omega_l)\right)+U.
\end{equation}
We calculate Green's function self-consistently with Dyson's equation,
\begin{equation}
G_{\Vvecs{k}}(\epsilon_n)^{-1}=i\epsilon_n+\mu-\epsilon^0_{\Vvecs{k}}-\Sigma_{\Vvecs{k}}(\epsilon_n).
\end{equation}
The irreducible particle-hole vertex $\Gamma^{\rm I}_{\Vvecs{kk'}}$
which satisfies $\Gamma^{\rm
I}_{\Vvecs{kk'}}(\epsilon_n,\epsilon_{n'})=\delta \Sigma_{\Vvecs{k}}(\epsilon_n)/\delta
G_{\Vvecs{k'}}(\epsilon_{n'})$ is given by
\begin{equation}
\Gamma^{\rm I}_{\Vvecs{kk'}}(\epsilon_n,\epsilon_{n'})=V_{\Vvecs{k}-\Vvecs{k'}}(\epsilon_n-\epsilon_{n'})\label{ward}
\end{equation}
where the Maki-Thompson term is taken into account and the Aslamazov-Larkin
term is omitted because the latter is negligible for the CVC.\cite{kontani-Hall}

In this paper we take $N=N_x\times N_y \times N_z=64\times 64\times 32$
$\Vvec{k}$-point meshes and the Matsubara frequencies $\epsilon_n$ takes
the value from
$-(2N_c-1)\pi T$ to $(2N_c-1)\pi T$ with $N_c=256$.
\subsection{conductivity}
In order to derive the transport coefficient, we begin with Kubo
formula,
\begin{equation}
\sigma_{\mu\nu}=e^2\sum_{{\Vvecs{kk'}}\sigma\sigma'}v^0_{{\Vvecs{k}}\mu} v^0_{{\Vvecs{k'}}\nu}\left.\frac{{\rm Im}K_{{\Vvecs{k}}\sigma,{\Vvecs{k'}}\sigma'}(\omega+i\delta)}{\omega}\right|_{\omega=0},
\end{equation}
where $K(\omega+i\delta)$ is analytic continuation of below $K(i\omega)$
\begin{eqnarray}
K_{{\Vvecs{k}}\sigma,{\Vvecs{k'}}\sigma'}(i\omega_n)&=&\int^{\beta}_0d\tau e^{\omega_n\tau}\nonumber\\
&&\times\left\langle T_\tau \left\{c^\dagger_{{\Vvecs{k}}\sigma}(\tau)c_{{\Vvecs{k}}\sigma}(\tau)c^\dagger_{{\Vvecs{k'}}\sigma'}c_{{\Vvecs{k'}}\sigma'}\right\}\right\rangle,
\end{eqnarray}
and
\begin{equation}
v^0_{{\Vvecs{k}}\mu}=\frac{\partial\epsilon^0_{\Vvecs{k}}}{\partial k_\mu}.
\end{equation}
Eliashberg\cite{elia} derived the conductivity in this way.
By generalizing Eliashberg's theory\cite{kontani-kino,kontani-Hall} the
conductivity is obtained as 
\begin{eqnarray}
\sigma_{xx}&=&e^2\sum_{\Vvecs{k}}\int\frac{d\epsilon}{\pi}\left(-\frac{\partial f}{\partial\epsilon}\right)\left\{v_{{\Vvecs{k}}x}(\epsilon)|G^{\rm R}_{\Vvecs{k}}(\epsilon)|^2J_{{\Vvecs{k}}x}(\epsilon)\right.\nonumber\\
&&\left.-{\rm Re}[G^{\rm R}_{\Vvecs{k}}(\epsilon)^2]\Vvec{v}_{\Vvecs{k}x}(\epsilon)^2\right\},\label{cond}
\end{eqnarray}
\begin{equation}
\Vvec{v}_{\Vvecs{k}}(\epsilon)=\Vvec{\nabla}[\epsilon^0_{\Vvecs{k}}+{\rm Re}\Sigma_{\Vvecs{k}}(\epsilon)],
\end{equation}

where $\Vvec{v}_{\Vvecs{k}}$ is the quasi-particle velocity (without
$Z$-factor) and the retarded Green's function $G^{\rm R}_{\Vvecs{k}}(\epsilon)$ is
derived by the analytic continuation. The total current $\Vvec{J}_{\Vvecs{k}}$ is given by the Bethe-Salpeter equation
\begin{equation}
\Vvec{J}_{\Vvecs{k}}(\epsilon)=\Vvec{v}_{\Vvecs{k}}(\epsilon)+\sum_{\Vvecs{k'}}\int^\infty_{-\infty}\frac{d\epsilon'}{4\pi i}{\cal T}_{22}^{\rm I}(\Vvec{k}\epsilon,\Vvec{k'}\epsilon')|G^{\rm R}_{\Vvecs{k'}}(\epsilon')|^2\Vvec{J}_{\Vvecs{k'}}(\epsilon'),\label{bethe}
\end{equation}
which is based on the Ward identity.
The irreducible four-point vertex ${\cal
T}^{\rm I}_{lm}(\Vvec{k}\epsilon,\Vvec{k}'\epsilon')$ is defined in
Ref. \onlinecite{elia}. According to eq. (\ref{ward}), it is given by
\begin{eqnarray}
{\cal T}_{22}^{\rm I}(\Vvec{k}\epsilon,\Vvec{k}'\epsilon')&=&\left({\rm cotanh}\frac{\epsilon'-\epsilon}{2T}-{\rm tanh}\frac{\epsilon'}{2T}\right)\nonumber\\
&&\times 2i{\rm Im}V_{\Vvecs{k'}-\Vvecs{k}}(\epsilon'-\epsilon+i\delta).
\end{eqnarray}

\subsection{Hall coefficient}
When a weak magnetic field $B$ is induced along $z$ axis, the Hall coefficient $R_H$ is given by
\begin{equation}
R_H=\frac{\sigma_{xy}/B}{\sigma_{xx}\sigma_{yy}}.
\end{equation}
The Hall conductivity $\sigma_{xy}$ is obtained by\cite{kohno,kontani-Hall}
\begin{eqnarray}
\sigma_{xy}/B&=&-e^3\sum_{\Vvecs{k}}\int\frac{d\epsilon}{2\pi}\left(-\frac{\partial f}{\partial\epsilon}\right)\nonumber\\
&&|{\rm Im}G^{\rm R}_{\Vvecs{k}}(\epsilon)||G^{\rm R}_{\Vvecs{k}}(\epsilon)|^2A_{\Vvecs{k}}(\epsilon),\\
A_{\Vvecs{k}}(\epsilon)&=&v_{\Vvecs{k}x}(\epsilon)\left[J_{\Vvecs{k}x}(\epsilon)\frac{\partial J_{\Vvecs{k}y}(\epsilon)}{\partial k_y}-J_{\Vvecs{k}y}(\epsilon)\frac{\partial J_{\Vvecs{k}x}(\epsilon)}{\partial k_y}\right]\nonumber\\
&&+\langle x\leftrightarrow y\rangle,\label{A}
%z_{\Vvecs{k}}&=&\left[1-\frac{\partial{\rm Re}\Sigma_{\Vvecs{k}}(\omega)}{\partial\omega}\right]^{-1}\\
%gamma_{\Vvecs{k}}(\epsilon)&=&-{\rm Im}\Sigma_{\Vvecs{k}}(\epsilon),
\end{eqnarray}
where $f(\epsilon)=1/(\exp(\epsilon/T)+1)$.
\subsection{Nernst effect}
Nernst coefficient under a weak magnetic field $B$ along $z$ axis and
gradient of temperature along $x$ axis is defined as
\begin{equation}
\nu=\frac{-E_y}{B\partial_x T}.
\end{equation}
According to the linear response theory\cite{mahan}, the response function
$L^{21}_{\mu\nu}$ is defined as
\begin{equation}
L^{21}_{\mu\nu}(i\omega_l)=-\frac{T}{\omega_l}\int^\beta_0 d\tau e^{i\omega\tau}
\langle T_\tau j^Q_\mu(\tau)j_\nu(\tau=0)\rangle.\label{lin}
\end{equation}
The electron current operator $\Vvec{j}$ and the heat current operator
$\Vvec{j}^Q$ are given by
\begin{equation}
\Vvec{j}=e\sum_{\Vvecs{k},\sigma}\Vvec{v}^0_{\Vvecs{k}}c^\dagger_{\Vvecs{k},\sigma}c_{\Vvecs{k},\sigma},
\end{equation}
and
{\small
\begin{eqnarray}
\Vvec{j}^Q(\omega_l)&=&\lim_{\tau'\rightarrow\tau}\sum_{\Vvecs{k},\sigma}\int^\beta_0 d\tau e^{i\omega_l\tau}\nonumber\\
&&\frac{1}{2}\left(\frac{\partial}{\partial\tau}-\frac{\partial}{\partial\tau'}\right)
c^\dagger_{\Vvecs{k},\sigma}(\tau)c_{\Vvecs{k},\sigma}(\tau')\\
&=&\frac{T}{N}\sum_{\Vvecs{k},n,\sigma}i(\epsilon_n+\omega_l/2)\Vvec{v}^0_{\Vvecs{k}}c^\dagger_{\Vvecs{k},\sigma}(\epsilon_n)c_{\Vvecs{k},\sigma}(\epsilon_n+\omega_l),
\end{eqnarray}
}
respectively.

After the analytic continuation for
eq. (\ref{lin}), $L^{21}_{xx}$ and $L^{21}_{xy}$
are obtained as\cite{kontani-thermo}
{\small
\begin{eqnarray}
L^{21}_{xx}(+i\delta)&=&eT\sum_{\Vvecs{k}}\int\frac{d\epsilon}{\pi}\left(-\frac{\partial f}{\partial\epsilon}\right)\Vvec{q}_{\Vvecs{k}x}(\epsilon)\nonumber\\
&&\times\left\{|G^{\rm R}_{\Vvecs{k}}(\epsilon)|^2J_{\Vvecs{k}x}(\epsilon)
-{\rm Re}[G^{\rm R}_{\Vvecs{k}}(\epsilon)^2]\Vvec{v}_{\Vvecs{k}x}(\epsilon)\right\},
\end{eqnarray}
}
{\small
\begin{eqnarray}
L^{21}_{xy}(+i\delta)&=&BTe^2\sum_{\Vvecs{k}}\int\frac{d\epsilon}{\pi}\left(-\frac{\partial f}{\partial\epsilon}\right)\nonumber\\
&&\times|{\rm Im}G^{\rm R}_{\Vvecs{k}}(\epsilon)||G^{\rm R}_{\Vvecs{k}}(\epsilon)|^2A'_{\Vvecs{k}}(\epsilon),\\
\gamma_{\Vvecs{k}}(\epsilon)&=&-{\rm Im}\Sigma_{\Vvecs{k}}(\epsilon),\\
A'_{\Vvecs{k}}(\epsilon)&=&\gamma_{\Vvecs{k}}(\epsilon)\left\{\Vvec{Q}_{\Vvecs{k}}(\epsilon)\times[\Vvec{v}_{\Vvecs{k}}(\epsilon)\times\Vvec{\nabla}]_z\left[\frac{\Vvec{J}_{\Vvecs{k}}(\epsilon)}{\gamma_{\Vvecs{k}}(\epsilon)}\right]\right\}_z,\label{A'}\\
\Vvec{Q}_{\Vvecs{k}}(\epsilon)&=&\Vvec{q}_{\Vvecs{k}}(\epsilon)+\sum_{\Vvecs{k}'}\int\frac{d\epsilon'}{4\pi i}\nonumber\\
&&\times{\cal T}_{22}^{\rm I}(\Vvec{k}\epsilon,\Vvec{k}'\epsilon')|G^{\rm R}_{\Vvecs{k}'}(\epsilon')|^2\Vvec{Q}_{\Vvecs{k}'}(\epsilon'),
\end{eqnarray}
}
where $\Vvec{q}_{\Vvecs{k}}$ is the quasi-particle heat velocity
$\Vvec{q}_{\Vvecs{k}}(\epsilon)=\epsilon\Vvec{v}_{\Vvecs{k}}(\epsilon)$. 

Using above expressions, $\nu$ can be rewritten as
\begin{equation}
\nu=\frac{L^{21}_{xy}}{BT^2\sigma_{xx}}-\frac{S\sigma_{xy}}{B\sigma_{xx}},
\end{equation}
where the thermopower $S$ is given by
\begin{equation}
S=\frac{1}{T^2}\frac{L^{21}_{xx}}{\sigma_{xx}}.
\end{equation}
\section{Result}
\begin{figure}[htdp]
\begin{center}
\includegraphics[height=55mm]{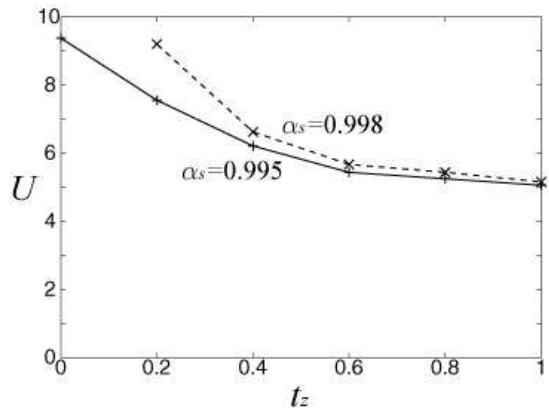}
\caption{$U$ against $t_z$ for $\alpha_S=0.995$ (solid line) and
 $\alpha_S=0.998$ (dotted line) at $T=0.02$.}
\label{st}
\end{center}
\end{figure}

Here, we show numerical results obtained by the CVC-FLEX approximation.
We use filling $n=0.9$ ($n=1$ corresponds to half filling) and the
intralayer hopping parameters $t_1=1$, $t_2=-1/6$ and $t_3=1/5$
which reproduce the Fermi surface of 2D high-$T_c$ cuprate YBCO, and
introduce the interlayer hopping $t_z$ which makes the Fermi surface
three-dimensional. The Stoner factor $\alpha_S={\rm
max}\left\{U\chi_{\Vvecs{q}}^0(\omega=0)\right\}$ represents the ``distance''
from the AF order ($\alpha_S=1$ corresponds to the boundary of the AF or
the spin density wave (SDW) order) since the
denominator of the static spin susceptibility is $1-U\chi^0$. 
%Because $\alpha_S$ depends on $t_z$, 
We calculate for each $t_z$ with keeping $\alpha_S$, by tuning
the value of $U$ as shown by solid line for $\alpha_S=0.995$ and
dotted line for $\alpha_S=0.998$ at $T=0.02$ in Fig. \ref{st}.
The ``distance'' from the AF order is considered to be same along these lines.
Note that $\alpha_S<1$ is always satisfied in 2D ($t_z=0$) at finite
temperatures reflecting the theory of
Mermin-Wagner.\cite{mermin-wagner}

\begin{figure}[htdp]
\begin{center}
\includegraphics[height=90mm]{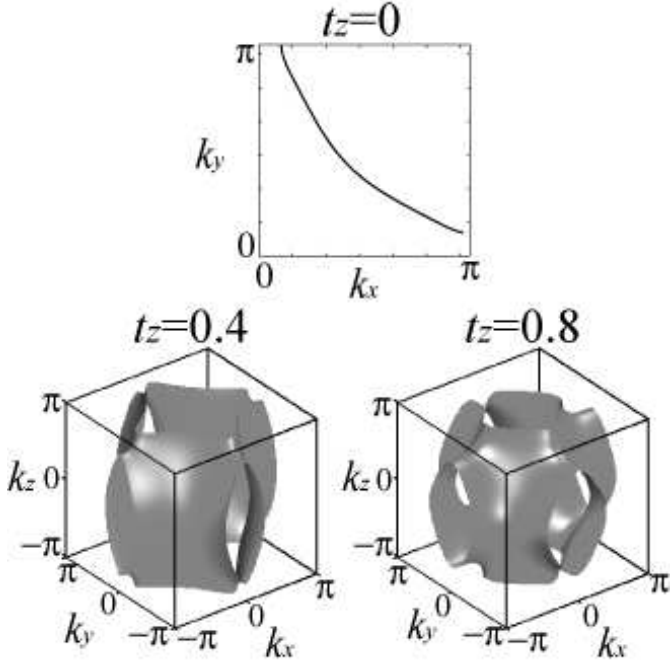}
\caption{Fermi surface for 2D with $t_z=0$ (upper panel), q3D with $t_z=0.4$ (lower left
 panel) and 3D with $t_z=0.8$ (lower right panel) for $T=0.02$, $\alpha_S=0.995$.}
\label{fermi}
\end{center}
\end{figure}

\begin{figure}[htdp]
\begin{center}
\includegraphics[height=25mm]{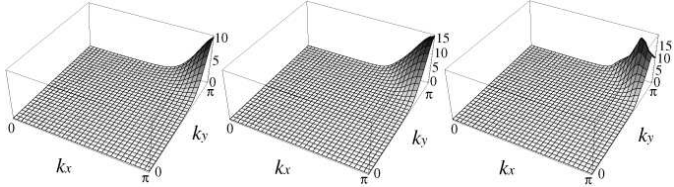}
\caption{The spin susceptibility for $t_z=0$ (left panel), $0.4$ (middle
 panel) with $k_z=\pi$, $0.8$ (right panel) with $k_z=\pi$
 , for $\alpha_S=0.995$, $T=0.02$.}
\label{chi}
\end{center}
\end{figure}

\begin{figure}[htdp]
\begin{center}
\includegraphics[height=55mm]{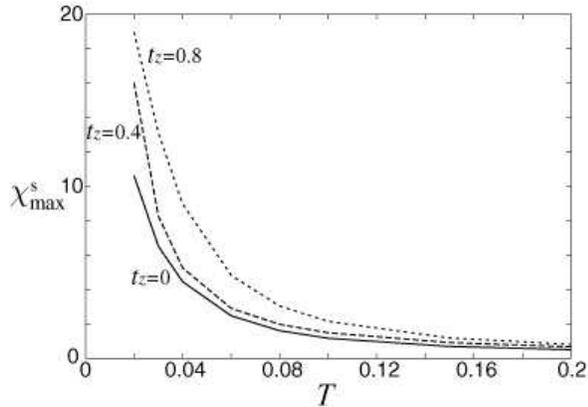}
\caption{The temperature dependence of the peak value of spin susceptibility for $t_z=0$, $0.4$,
 $0.8$, in the condition of $\alpha_S=0.995$ at $T=0.02$.}
\label{chizz3}
\end{center}
\end{figure}

Fig. \ref{fermi} shows the Fermi surface for $t_z=0$ (2D), $t_z=0.4$
(quasi-3D; q3D) and $t_z=0.8$ (3D). We see that three dimensionality becomes
stronger as the value of $t_z$ increases. 
The $\Vvec{k}$-dependence of the static spin susceptibility $\chi^{\rm s}_{\Vvecs{k}}$ is
shown in Fig. \ref{chi}, where the peak position is commensurate; $(k_x,k_y)=(\pi,\pi)$
for $t_z=0$ (2D), $(k_x,k_y,k_z)=(\pi,\pi,\pi)$ for $t_z=0.4$ (q3D) and
incommensurate around $(\pi,\pi,\pi)$ for $t_z=0.8$ (3D). From these
peak structures, we ensure that the AF
fluctuations are dominant in these systems.
In this case, the peak values of $\chi^{\rm s}$ increase with $t_z$ as
seen in Fig. \ref{chizz3}. 
This means that the present system approaches to the AF instability as
the dimensionality changes from 2D to 3D.

\begin{figure}[htdp]
\begin{center}
\includegraphics[height=90mm]{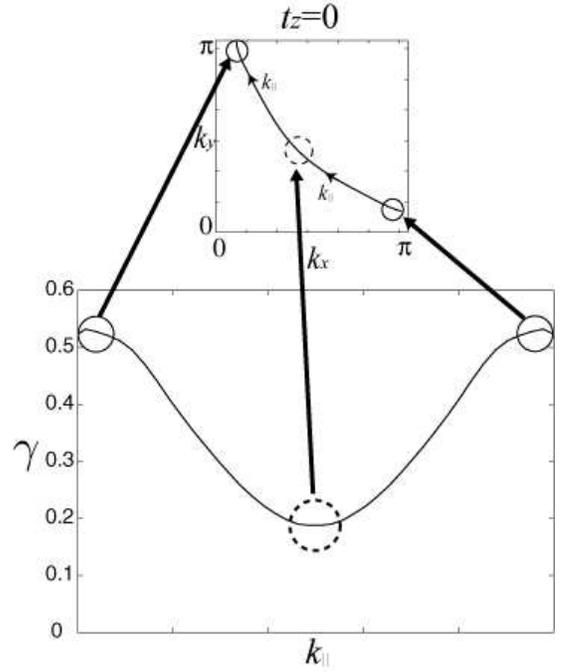}
\caption{$\gamma$ along the Fermi surface for $t_z=0$, $\alpha_S=0.995$ at $T=0.02$, where the hot spots and cold spots are depicted by
 solid circles and dotted circle, respectively.}
\label{del-tz0}
\end{center}
\end{figure}

\begin{figure}[htdp]
\begin{center}
\includegraphics[height=100mm]{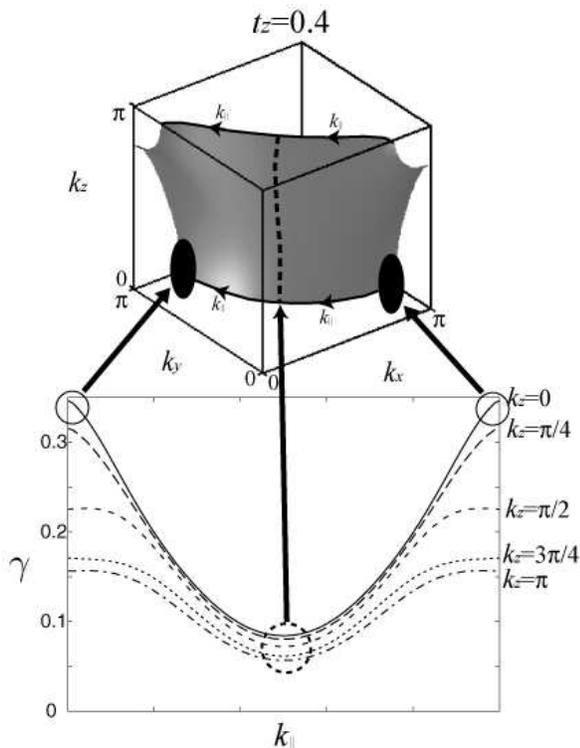}
\caption{$\gamma$ along the Fermi surface for $t_z=0.4$,
 $\alpha_S=0.995$ at $T=0.02$, where the hot spots and the ``cold lines'' are depicted
 by circles and dotted thick line, respectively. The trajectories of
 $k_\parallel$ are depicted on the Fermi surface in $k_z=0,\pi$ as examples.}
\label{del-tz0.4}
\end{center}
\end{figure}

\begin{figure}[htdp]
\begin{center}
\includegraphics[height=100mm]{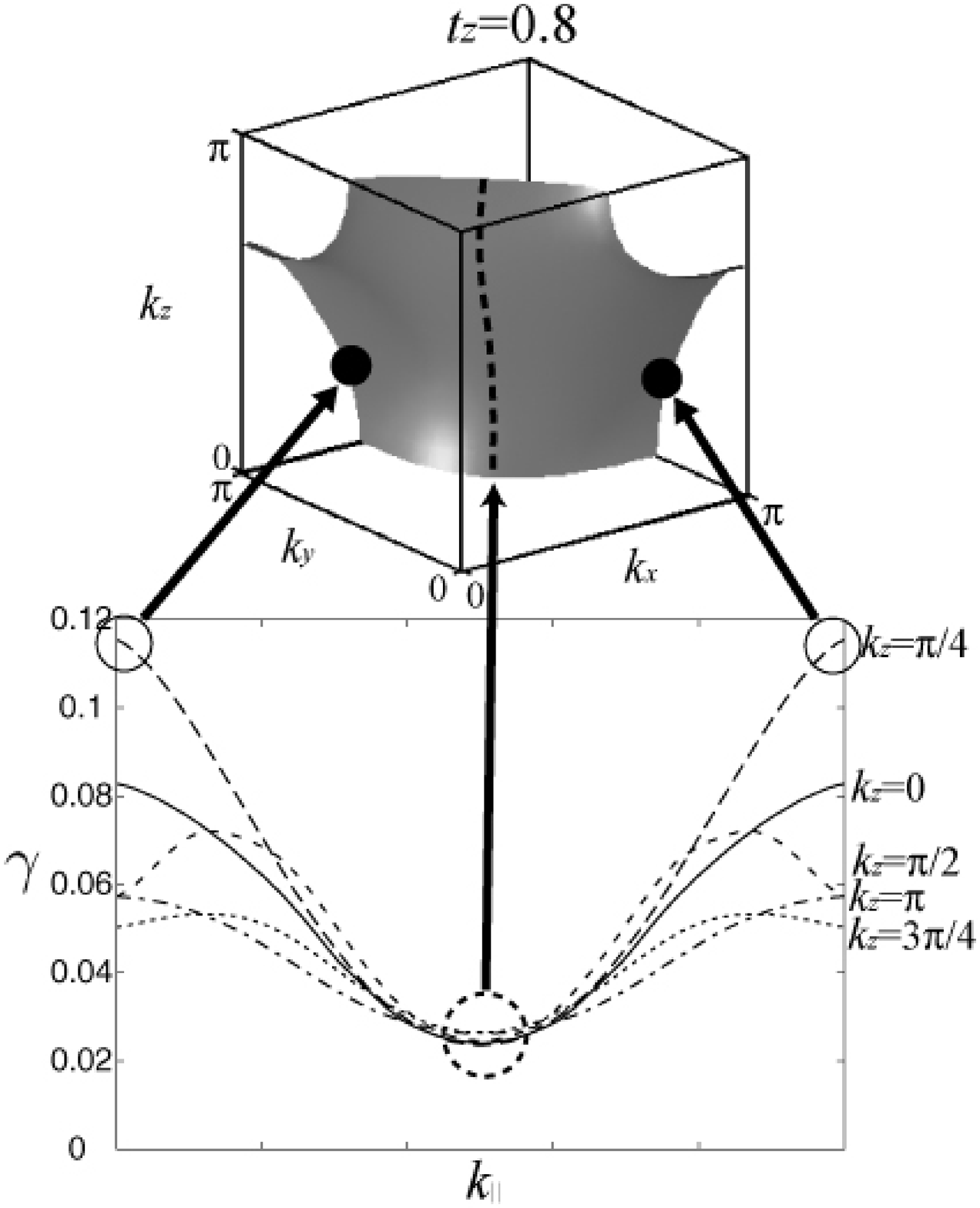}
\caption{$\gamma$ along the Fermi surface for $t_z=0.8$,
 $\alpha_S=0.995$, at $T=0.02$ where the hot spots and the ``cold
 lines'' are depicted by circles and dotted thick line, respectively.}
\label{del-tz0.8}
\end{center}
\end{figure}

To understand 3D structure of the relaxation time $\tau$, we show the
$\Vvec{k}$-dependence of $\gamma_{\Vvecs{k}}(=-{\rm
Im}\Sigma_{\Vvecs{k}}(\epsilon=0))\propto 1/\tau_{\Vvecs{k}}$ along the Fermi
surface in Figs. \ref{del-tz0}-\ref{del-tz0.8}, where hot spots are depicted by
circles and cold spots are illustrated by dotted circle. The bottom
panels represent the momentum dependence of $\gamma_{\Vvecs{k}}$ along
the Fermi surface for each $k_z$.
We see the hot spots exist on the plains of $k_z=0$ and $k_z=\pi/4$ for $t_z=0.4$ and
$t_z=0.8$, respectively.  

The volume fraction of hot spots decreases as $t_z$ increases, and for
three-dimensional case ($t_z=0.8$) hot spots have point-like
structure (``hot points''), which is consistent with the experimental
result obtained by de Haas-van Alphen measurements on CeIn$_3$.\cite{ebihara}
Generally in 3D systems, nesting exists in only small parts of the Fermi
surface, and hot spots form there.
 In general, the ``hot lines''\cite{rosch} where hot spots form
line-like structure would not be appropriate in 3D systems.

On the other hand, $\gamma_k$ increases more sharply along $k_\parallel$ direction than $k_z$ direction around minimum point of $\gamma_k$ for
$t_z=0.4$ and $0.8$ as depicted in Fig. \ref{del-tz0.4} and
Fig. \ref{del-tz0.8}. In this sense, cold spots stretch strikingly along
$k_z$ direction. Then, cold spots form line-like structure (``cold
lines'').
They are
aligned perpendicular to the plain with hot spots. The formation of hot
points and cold lines would be generally expected in three-dimensional
systems.
To confirm the generality, we must study much more systems with various types of
3D Fermi surfaces.



\begin{figure}[htdp]
\begin{center}
\includegraphics[height=55mm]{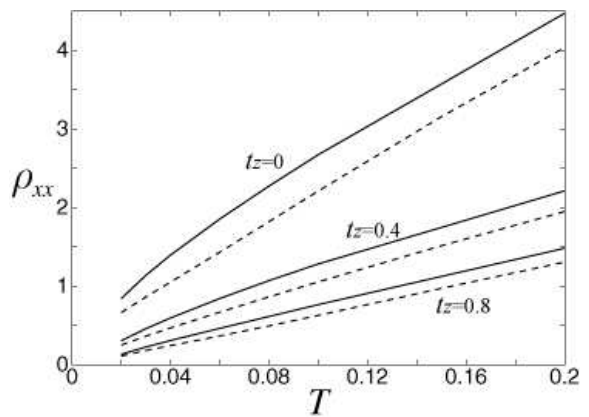}
\caption{The resistivity $\rho$ with the CVC (solid line) and
 without the CVC (dotted line) for $t_z=0$, $0.4$,
 $0.8$ as a function of $T$ with $\alpha_S=0.995$ at $T=0.02$.}
\label{rho3}
\end{center}
\end{figure}

The calculated temperature dependence of the resistivity
$\rho=1/\sigma_{xx}$ for $t_z=0$ (2D), $t_z=0.4$ (q3D) and $t_z=0.8$
(3D) with
$U=9.4$, $U=6.2$ and $U=5.2$, respectively (see Fig. \ref{st}), are
shown in Fig. \ref{rho3}, where the unit of $\rho$ is $\frac{\hbar
a_c}{e^2}\sim 2.4\times 10^{-6}\Omega \rm{m}$ for a bilayer YBCO (lattice constant along $c$ axis : $a_c=5.8\times 10^{-10}$m) and
$3.1\times 10^{-6}\Omega \rm{m}$ for CeCoIn$_5$ ($a_c=7.6\times 10^{-10}$m),
respectively.
The value of $U$ is chosen to satisfy $\alpha_S=0.995$ at $T=0.02$ for each case.
In this case, the ``distance'' from the AF QCP is the same among three
$t_z$ parameters.  We see that,
independently of the dimensionality, the
resistivities with and without the CVC are proportional to the
temperature. Then, the value of $\rho$ is slightly enhanced by the CVC.
It decreases as $t_z$ increases since the corresponding value of
$U$ is reduced. 
This $T$-liner behavior of the resistivity is consistent with experiments for the
systems close to the AF QCP, such as two-dimensional high-$T_c$ 
cuprate and three-dimensional CeCoIn$_5$\cite{nakajima}. 
In detail, the resistivities with the CVC show sub-liner temperature
dependence in low temperatures, which are also observed in the
experimental results for heavy fermion CeRhIn$_5$ (private discussion).
According to the SCR theory,\cite{moriya-ueda,moriya-takimoto} the
resistivity behaves as $\rho\propto T$ in 2D,
and $\rho\propto T^{3/2}$ in 3D.
However, the SCR theory also predicts that 
$\rho\propto T$ for a wide range of temperatures
even in 3D systems when the system is close to the AF QCP,
which is consistent with the present numerical calculation.

\begin{figure}[htdp]
\begin{center}
\includegraphics[height=55mm]{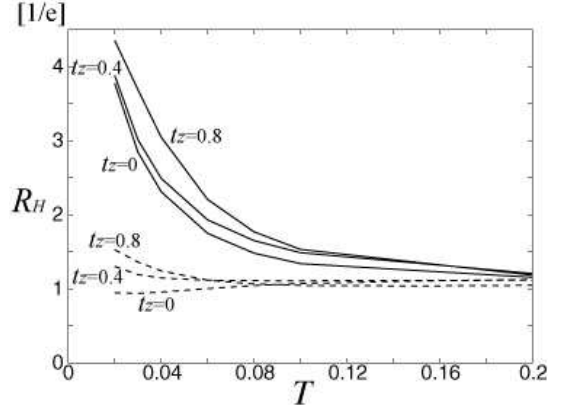}
\caption{The Hall coefficient $R_H$ with the CVC (solid line) and
 without the CVC (dotted line) for $t_z=0$, $0.4$,
 $0.8$ as a function of $T$, with $\alpha_S=0.995$ at $T=0.02$.}
\label{h3}
\end{center}
\end{figure}

Next, we show the temperature dependence of the Hall coefficient $R_H$
in Fig. \ref{h3}, where the unit of $R_H$ is $\frac{a_aa_ba_c}{e}\sim
5.2\times 10^{-10}\rm{m}^3/\rm{C}$ for a bilayer
YBCO (lattice constant along $a$ and $b$ axis : $a_a=a_b=3.8\times
10^{-10}$m and that along $c$ axis : $a_c=5.8\times 10^{-10}$m) and
$1.0\times 10^{-9}\rm{m}^3/\rm{C}$ for CeCoIn$_5$ ($a_a=a_b=4.6\times 10^{-10}$m
and $a_c=7.6\times 10^{-10}$m), respectively.
In regard to the horizontal axis, unit of the temperature $t_1$ is $4000$K
for YBCO. On the other hand, we estimate the nearest neighbor hopping $t_1=400$K for CeCoIn$_5$, because the
experimetal data\cite{nakajima} shows that magnitude of $R_H$ begins to increase below $40$K, which
corresponds to $0.1t_1$ in Fig. \ref{h3}.
The $R_H$ without the CVC, which corresponds to the RTA, is almost constant. However, independently of the dimensionality,
 $R_H$ with the CVC increases as temperature decreases, which is consistent with experimental results
 for high-$T_c$ cuprate\cite{takeda} and heavy fermion compounds (CeCoIn$_5$\cite{nakajima},
 CeCu$_{6-x}$Au$_x$\cite{fukuhara} and YbRh$_2$Si$_2$\cite{paschen}).
Namely, the RTA cannot explain the strong temperature dependence of the
Hall coefficient close to
 the AF QCP. Moreover, in 3D case hot spots take point-like shape, which
 means that the effective electron density for transport phenomena
 ($n_{\rm eff}$) is large compared with two-dimensional case.
Since $R_H^{\rm RTA}\approx\frac{1}{en_{\rm eff}}$ is satisfied in the
RTA, $R_H^{\rm eff}$
cannot become large in 3D systems.

%In 2D system, Stojkovi\'c and Pines\cite{stojkovic} derived an
%approximate expression based on the RTA for $\tau_{\rm
%cold}/\tau_{\rm hot}\gg 1$,
%\begin{equation}
%R_H^{\rm RTA}\approx\frac{1}{2ne}\sqrt{\frac{\tau_{\rm cold}}{\tau_{\rm hot}}\frac{m_{\rm hot}}{m_{\rm cold}}}.
%\label{sto}
%\end{equation}
%In the present result, the effective mass satisfy following relation
%$m_{\rm hot}\approx m_{\rm cold}$, and we can estimate $\tau_{\rm
%cold}/\tau_{\rm hot}\approx 2.6$, $6.0$, $5.1$ for $t_z=0$, $0.4$, $0.8$,
%respectively in Fig. \ref{del-tz0}-\ref{del-tz0.8}.
%$R_H$ for $t_z=0.8$ is approximately 1.5 times larger than that for $t_z=0$.
%However, eq. (\ref{sto}) gives an overestimated value of $R_H^{\rm RTA}$ in 3D
%system because the anisotropy of $\tau$ for $k_z$ direction is neglected. In
%the $k_xk_y$-plane which does not include hot spots, $\tau_{\rm max}/\tau_{\rm min}$ is quite
%smaller than that in the plane including hot spots. Using eq. (\ref{sto}) in 3D
%corresponds to assuming $\tau_{\rm max}/\tau_{\rm min}$=$\tau_{\rm
%cold}/\tau_{\rm hot}$ for all plane, which obviously
%overestimates. Thus, $R_H^{\rm RTA}({\rm 3D})$ should be smaller than
%the value given by eq. (\ref{sto}).

As a result, the CVC is indispensable to explain the behavior
 of $R_H$ in 3D close to the AF QCP. In the present results, the
 maximum enhancement of $R_H$ is given by $R_H(T=0.02)/R_H(t=0.2)\sim 5$
 for $t_z=0.8$ and $\alpha_S=0.998$. $R_H$ should increase further if we
 calculate at lower temperatures. The shape of the Fermi surface in CeCoIn$_5$
 resembles to that of our model for $t_z=0.8$. However, to reproduce the experimental results
 in CeCoIn$_5$ quantitatively, we have to study it based on the
 realistic band structures of CeCoIn$_5$.\cite{maehira,harima}
Our calculation shows that $R_H$ is strongly
enhanced by the CVC even in 3D systems, and its maximum value becomes as
large as that in 2D. 

Here, we discuss the reason why $R_H$ is strongly enhanced by the CVC in 3D systems
based on the numerical study.
The general expression for $\sigma_{xy}$ is given
by\cite{kohno,fukuyama}
\begin{equation}
 \sigma_{xy}/B=-\frac{e^3}{8\pi}\int dk_z\oint_{\rm FS}dk_\parallel|\Vvec{J}_{\Vvecs{k}}|_\perp^2\left(\frac{\partial\theta_{\Vvecs{J}}(\Vvec{k})}{\partial k_\parallel}\right)\frac{1}{(\gamma_{\Vvecs{k}})^2},\label{sigma}
\end{equation}
where $k_{\parallel}$ is the component of $\Vvec{k}$ along the unit
vector
$\Vvec{e}_{\parallel}(\Vvec{k})=(\Vvec{e}_z\times\Vvec{v}_{\Vvecs{k}})/|\Vvec{v}_{\Vvecs{k}}|$
which is in the $k_xk_y$-plane and parallel to the Fermi surface.
$|\Vvec{J}_{\Vvecs{k}}|_\perp=\sqrt{J_{\Vvecs{k}x}^2+J_{\Vvecs{k}y}^2}$,
and $\theta_{\Vvecs{J}}$ is the angle between the total
current $\Vvec{J}$ and the $x$ axis.
In this line integration $\Vvec{k}$-point moves anticlockwise along the
Fermi surface around the $k_z$ axis.

\begin{figure}[htdp]
\begin{center}
\includegraphics[height=55mm]{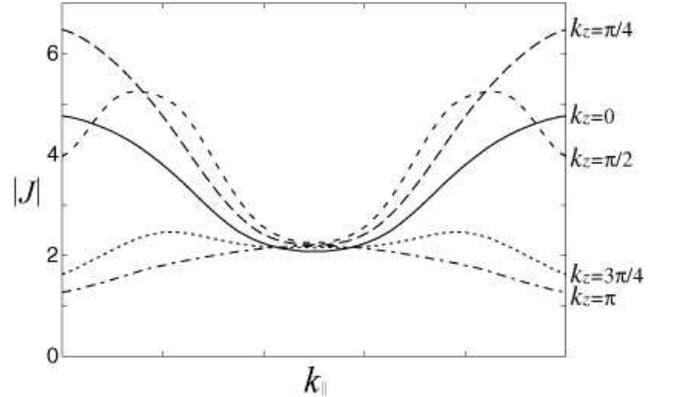}
\caption{Absolute value of the total current $\Vvec{J}$ with the CVC
 along the Fermi surface for $t_z=0.8$ and $t_z=0$ (thin line) with $\alpha_S=0.995$ at $T=0.02$. }
\label{abs-tz0.8}
\end{center}
\end{figure}

We see that for $t_z=0.8$, cold spots (where $\gamma_{\Vvecs{k}}$ is
small) form lines (cold lines) at the center of the
Fermi surface along the $z$ axis in
Fig. \ref{del-tz0.8}. 
%In this wide cold region, absolute values of $\Vvec{J}$
%for $k_z=\pi/4,\pi/2$ are larger than that for other $k_z$ as seen in
%Fig. \ref{abs-tz0.8}. 
As shown in the last term of eq. (\ref{sigma}), main contribution for $\sigma_{xy}$ is
expected to come from the cold lines. 
%We see that in Fig. \ref{abs-tz0.8},
%$|\Vvec{J}_{\Vvecs{k}}|_\perp$ shows similar behavior to
%$\gamma_{\Vvecs{k}}$ in Fig. \ref{del-tz0.8}. Then, in eq. (\ref{sigma})
%$\partial\theta_{\Vvecs{J}}(\Vvec{k})/{\partial k_\parallel}$ plays
%more important role for $\sigma_{xy}$ than
%$|\Vvec{J}_{\Vvecs{k}}|_\perp/\gamma_{\Vvecs{k}}$ 
We see that in Fig. \ref{abs-tz0.8} the momentum dependence of the absolute
value of the total current $|\Vvec{J}_{\Vvecs{k}}|$ is quite similar to that of
$\gamma_{\Vvecs{k}}$. 
%However, the momentum dependence of $\gamma_{\Vvecs{k}}$
%is little prominent as compared with that of $|\Vvec{J}_{\Vvecs{k}}|$.

\begin{figure}[htdp]
\begin{center}
\includegraphics[height=55mm]{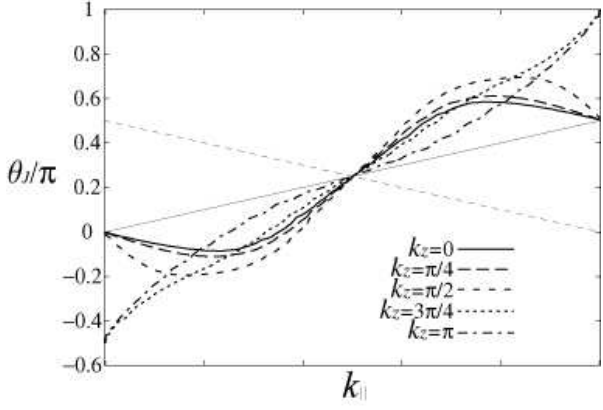}
\caption{the angle $\theta_{\Vvecs{J}}$ of the total current $\Vvec{J}$ with the CVC
 along the Fermi surface for $t_z=0.8$,
 $\alpha_S=0.995$ at $T=0.02$. Thin line and dotted thin line approximately
 correspond to the angle $\theta_{\Vvecs{v}}$ of the current without the
 CVC for $k_z=0$-$\pi/2$ and $k_z=3\pi/4$-$\pi$, respectively} 
\label{ang-tz0.8}
\end{center}
\end{figure}

In Fig. 12, we plot $\theta_{J}$ along the trajectories of 
$k_{\parallel}$.
As references, we also plot the corresponding quantity 
without the CVC, $\theta_{\Vvecs{v}}$ for $k_z=0$-$\pi/2$ and $k_z=3\pi/4$-$\pi$ as a thin line and a dotted thin line, respectively. 
As show in Fig. \ref{ang-tz0.8},
$\theta_{\Vvecs{v}}$ for $k_z=0$-$\pi/2$ and $k_z=3\pi/4$-$\pi$ decrease
and increase, respectively, along the trajectories of $k_\parallel$.
On the other hand, $\theta_{\Vvecs{J}}$ has a non-monotonic change along
these trajectories. Especially, for $k_z=3\pi/4$-$\pi$
$\theta_{\Vvecs{J}}$ decreases   contrary to the case of $\theta_{v}$.
We stress that the magnitude of $\partial\theta_{\Vvec{J}}/\partial k_{\parallel}$ becomes
larger than that without the CVC ($\partial\theta_{\Vvec{v}}/\partial k_{\parallel}$)around the cold lines. 

\begin{figure}[htdp]
\begin{center}
\includegraphics[height=55mm]{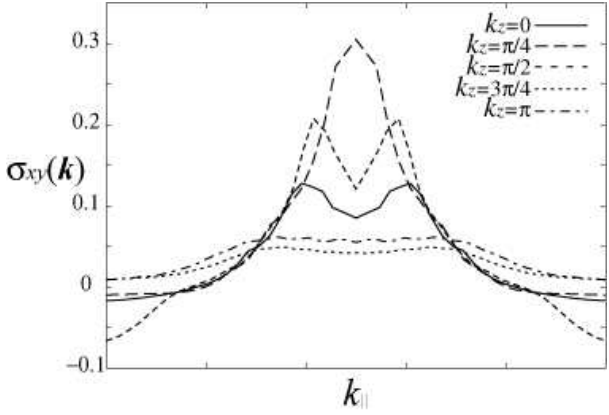}
\caption{$\sigma_{xy}(\Vvec{k})$ along the Fermi surface for $t_z=0.8$,
 $\alpha_S=0.995$ at $T=0.02$. }
\label{DHall-tz0.8}
\end{center}
\end{figure}

In Fig. \ref{DHall-tz0.8}, we plot the momentum resolved 
Hall conductivity $\sigma_{xy}(\Vvec{k})$ defined by 
$\sigma_{xy}=\sum_{\Vvecs{k}}\sigma_{xy}(\Vvec{k})$, 
where  $\sigma_{xy}(\Vvec{k})$ is given by 
$\sigma_{xy}(\Vvec{k})=-\int\frac{d\epsilon}{2\pi N}\left(-\frac{\partial
f}{\partial\epsilon}\right)|{\rm Im}G^{\rm
R}_{\Vvecs{k}}(\epsilon)||G^{\rm
R}_{\Vvecs{k}}(\epsilon)|^2A_{\Vvecs{k}}(\epsilon)$.
The magnitude of $\sigma_{xy}(\Vvec{k})$ 
takes large values around the cold lines 
, especially for $k_z=-\pi/2$-$\pi/2$. 
We should comment that for $k_z=3\pi/4,\pi$ the difference of the value of $\theta_{\Vvecs{v}}$ and 
$\theta_{\Vvecs{J}}$ becomes $\pi$ at the edge of the 
trajectory as shown in Fig. \ref{ang-tz0.8} due to the CVC.
At that time, 
the direction of
$\Vvec{J}_{\Vvecs{k}}$ is opposite to that of $\Vvec{v}_{\Vvecs{k}}$.
In this case, strong AF spin
fluctuation enhances the magnitude of ${\cal T}^{\rm
I}_{22}(\Vvec{k}\epsilon,\Vvec{k'}\epsilon')$ in the eq. (\ref{bethe}) for
$\Vvec{k}-\Vvec{k}'=\pm(\pi,\pi,\pi)$, $\pm(-\pi,\pi,\pi)$,
$\pm(\pi,-\pi,\pi)$ and $\pm(\pi,\pi,-\pi)$.
In this case, from  eq. (\ref{bethe}), we can obtain 
$\Vvec{J}_{\Vvecs{k}}\approx
\Vvec{v}_{\Vvecs{k}}+\alpha_{\Vvecs{k}}\Vvec{J}_{\Vvecs{k'}}$
($0<\alpha\leq 1$)  \cite{kontani-Hall}. 
This equation is easily solved as
\begin{equation}
\Vvec{J}_{\Vvecs{k}}=
\frac{\Vvec{v}_{\Vvecs{k}}+\alpha_{\Vvecs{k}}\Vvec{v}_{\Vvecs{k'}}}{1-\alpha_{\Vvecs{k}}\alpha_{\Vvecs{k'}}}\label{alpha}.
\end{equation}

\begin{figure}[htdp]
\begin{center}
\includegraphics[height=60mm]{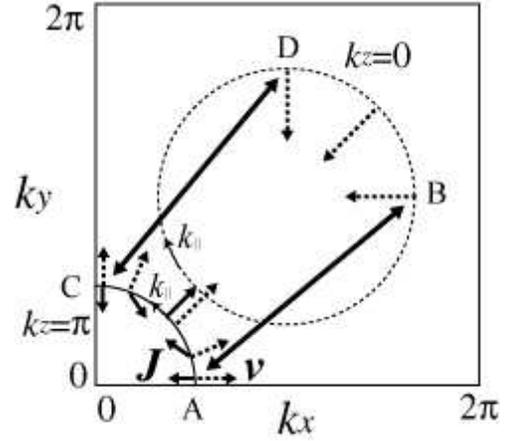}
\caption{Schematic illustration to show that at A and B $\Vvec{J}_{\Vvecs{k}}$ can
 direct opposite to $\Vvec{v}_{\Vvecs{k}}$, where the Fermi surface for
 $k_z=0,\pi$ are depicted by dotted and solid circle, respectively, and
 $\Vvec{J}_{\Vvecs{k}}$ and $\Vvec{v}_{\Vvecs{k}}$ are by solid and
 dotted arrows, and the thick left-right arrows represent the coupling points.}
\label{sche}
\end{center}
\end{figure}

In Fig. \ref{sche}, we illustrate schematic behaviors of the
quasi-particle velocity $\Vvec{v}_{\Vvecs{k}}$ and the total current
$\Vvec{J}_{\Vvecs{k}}$ on the Fermi surfaces sliced at $k_z=\pi$ (solid
circle) and at $k_z=0$ (dotted circle).
We focus on the position of points A and B, which are connected by
the nesting vector $\Vvec{k}-\Vvec{k}'=(-\pi,-\pi,\pi)$. Here, we write
the quasi-particle velocities at A
and B as $\Vvec{v}_{\Vvecs{k}}$ and $\Vvec{v}_{\Vvecs{k'}}$,
respectively.
We see that $\Vvec{v}_{\Vvecs{k}}$ and $\Vvec{v}_{\Vvecs{k'}}$ are
antiparallel and $|\Vvec{v}_{\Vvecs{k}}|<|\Vvec{v}_{\Vvecs{k'}}|$.
Considering that $\alpha\lesssim 1$, $\Vvec{J}_{\Vvecs{k}}$ given in
eq. (\ref{alpha}) takes an opposite direction of $\Vvec{v}_{\Vvecs{k}}$.
In the same way, the total current and the quasi-particle velocity at C are
also antiparallel.
This nontrivial behavior of $\Vvec{J}_{\Vvecs{k}}$ has not been pointed
out in previous studies for two-dimensional systems.
This feature might induce an anomalous transport phenomenon.

%In 2D case with a single Fermi surface, $\Vvec{J}_{\Vvecs{k}}$ can hardly have
%opposite direction to $\Vvec{v}_{\Vvecs{k}}$. Even if the
%direction of 
%$\Vvec{v}_{\Vvecs{k}}$ is opposite to $\Vvec{v}_{\Vvecs{k'}}$, both
%$\Vvec{v}_{\Vvecs{k}}$=$-\Vvec{v}_{\Vvecs{k'}}$ and
%$\alpha_{\Vvecs{k}}=\alpha_{\Vvecs{k'}}$ are satisfied when the main AF
%fluctuation is commensurate. From these equations and eq. (\ref{alpha})
%$\Vvec{J}_{\Vvecs{k}}=\Vvec{v}_{\Vvecs{k}}/(1+\alpha_{\Vvecs{k}})$ is obtained.
%It is peculiar to 3D systems that $\Vvec{J}_{\Vvecs{k}}$ can have
%opposite direction to $\Vvec{v}_{\Vvecs{k}}$ for usual
%single Fermi surface.

\begin{figure}[htdp]
\begin{center}
\includegraphics[height=55mm]{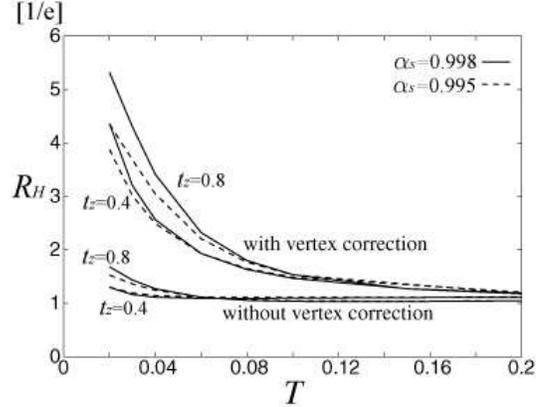}
\caption{The Hall coefficient $R_H$ with the CVC (upside) and
 without the CVC (downside) for $t_z=0.4$,
 $0.8$ as a function of $T$ with $\alpha_S=0.995$ (solid line),
 $0.998$ (dotted line) at $T=0.02$.}
\label{h-2}
\end{center}
\end{figure}

\begin{figure}[htdp]
\begin{center}
\includegraphics[height=58mm]{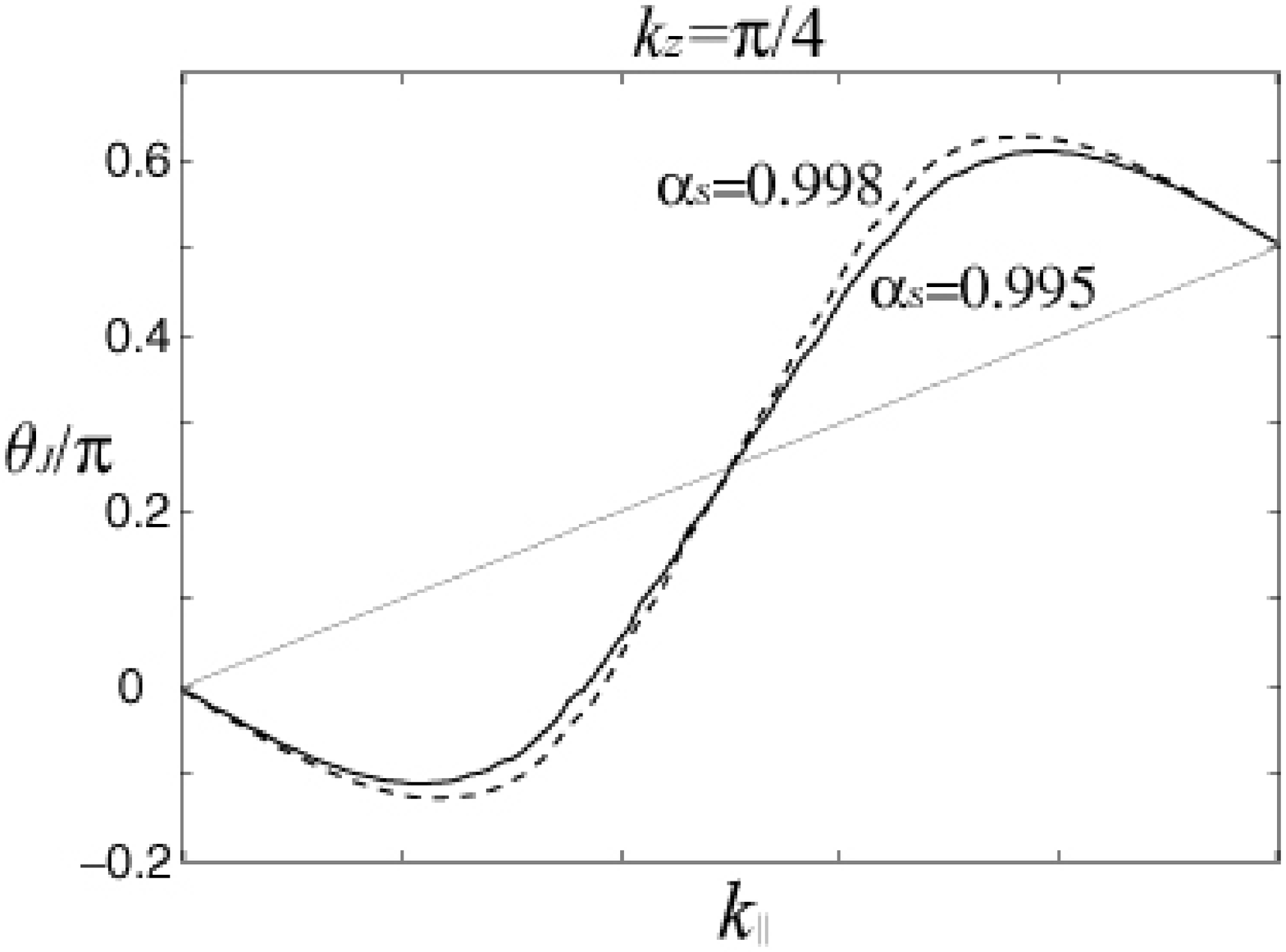}
\caption{the angle $\theta_{\Vvecs{J}}$ of the total current $\Vvec{J}$ with the CVC
 along the Fermi surface for $t_z=0.8$, $k_z=\pi/4$,
 $\alpha_S=0.995$ (solid line) and $\alpha_S=0.998$ (dotted line) at
 $T=0.02$. Thin line approximately corresponds to one without the CVC.}
\label{ang-tz0.8-st}
\end{center}
\end{figure}

In Fig. \ref{h-2}, we show the temperature dependence of $R_H$ 
 for $\alpha_S=0.998$ (solid line) and $\alpha_S=0.995$ (dotted
line) for $t_z=0.4$ and $0.8$. 
We see that $R_H$ with the CVC increases as $\alpha_S$ approaches unity. It
 seems that $R_H$ tends to diverge as the system approaches the AF QCP.
The reason can be understood by seeing Fig. \ref{ang-tz0.8-st}, where $\partial\theta_{\Vvecs{J}}/\partial
 k_\parallel$ takes a large value at the cold spot which corresponds to the center of the
 $k_\parallel$ axis.
According to eq. (\ref{sigma}), this fact leads to the strong enhancement of $\sigma_{xy}$. 

Finally, we discuss the Nernst coefficient $\nu$.
It is known that $\nu$ vanishes in a complete spherical system, which is called
Sondheimer cancellation.\cite{sondheimer,ong} Although this cancellation
is not perfect in real anisotropic systems, the magnitude of $\nu$ 
becomes small ($\sim 1{\rm nV/KT}$) in conventional metals.

However, $\nu$ is enhanced below $T^*$ (in the pseudo-gap region) for high-$T_c$ cuprates.
The authors of Refs. \onlinecite{ong2,ong} suggest a possibility that the vortex-like excitation emerges
in under-doped high-$T_c$ cuprates to explain the enhancement of $\nu$ in the pseudo-gap region.

On the other hand, one of the present
author\cite{kontani-nu,kontani-yamada} has shown that 
strong enhancement of $\nu$ for high-$T_c$ cuprates
is naturally derived based on the
FLEX+T-matrix approximation with the CVC.
Furthermore, CeCoIn$_5$ also shows huge {\it negative} $\nu$
below $20$K,\cite{bel} which cannot be ascribed to the vortex mechanism.
Here, we aim to reveal the mechanism of the unconventional
enhancement for $\nu$ close to the QCP irrespective of the dimensionality.

We estimate the renormalization factor ($z$) dependence of the transport
coefficients, before showing the result of $\nu$.
In the following, we will show that $\rho$ and $R_H$ are independent of $z$, and $\nu\propto z^{-1}$ in Hubbard model.
Using the relation\cite{kontani-thermo}
{\small
\begin{equation}
\sum_{\Vvecs{k}}=\frac{1}{(2\pi)^3}\int dS_{\Vvecs{k}}dk_\perp=\frac{1}{(2\pi)^3}\int \frac{dS_{\Vvecs{k}}d\epsilon^0_{\Vvecs{k}}}{|\Vvec{v}^0_{\Vvecs{k}}|}=\frac{1}{(2\pi)^3}\int \frac{dS_{\Vvecs{k}}d\epsilon^*_{\Vvecs{k}}}{z|\Vvec{v}_{\Vvecs{k}}|},
\end{equation}
}
where $S_{\Vvecs{k}}$ is the Fermi surface and $k_\perp$ represents the
momentum perpendicular to the Fermi surface and also using the relation $|G^{\rm
R}_{\Vvecs{k}}(\epsilon)|^2\sim \pi
z\delta(\epsilon-\epsilon^*_{\Vvecs{k}})/\gamma_{\Vvecs{k}}$
($\epsilon^*_{\Vvecs{k}}=z(\epsilon^0_{\Vvecs{k}}+\rm{Re}\Sigma_{\Vvecs{k}}(0)-\mu)$)
for
$\gamma\ll T$, we obtain
\begin{eqnarray}
\sigma_{xx}&\sim&\frac{e^2}{(2\pi)^3}\int \frac{dS_{\Vvecs{k}}d\epsilon^*_{\Vvecs{k}}}{z|\Vvec{v}_{\Vvecs{k}}|}\left(-\frac{\partial f}{\partial \epsilon}\right)_{\epsilon=\epsilon^*_{\Vvecs{k}}}\Vvec{v}_{\Vvecs{k}x}\Vvec{J}_{\Vvecs{k}x}\frac{z}{\gamma_{\Vvecs{k}}}\\
&\propto&(z)^0.
\end{eqnarray}

In the same way, $\sigma_{xy}$ is given by,
\begin{eqnarray}
\sigma_{xy}&\sim&\frac{e^3B}{4(2\pi)^3}\int \frac{dS_{\Vvecs{k}}d\epsilon^*_{\Vvecs{k}}}{z|\Vvec{v}_{\Vvecs{k}}|}\left(-\frac{\partial f}{\partial \epsilon}\right)_{\epsilon=\epsilon^*_{\Vvecs{k}}}A_{\Vvecs{k}}\frac{z}{\gamma_{\Vvecs{k}}^2}\\
&\propto&(z)^0.
\end{eqnarray}
We see that $\sigma_{xx}$ and $\sigma_{xy}$ are independent of $z$.
Thus, we confirm that $\rho=1/\sigma_{xx}\propto (z)^0$ and
$R_H=\frac{\sigma_{xy}}{B\sigma_{xx}^2}\propto (z)^0$ are independent of
$z$.
On the other hand, thermopower $S$ is given by
{\small
\begin{eqnarray}
S&\sim&\frac{e}{(2\pi)^3T\sigma_{xx}}\int \frac{dS_{\Vvecs{k}}d\epsilon^*_{\Vvecs{k}}}{z|\Vvec{v}_{\Vvecs{k}}|}\left(-\frac{\partial f}{\partial \epsilon}\right)_{\epsilon=\epsilon^*_{\Vvecs{k}}}\epsilon^*_{\Vvecs{k}}\Vvec{v}_{\Vvecs{k}x}\Vvec{J}_{\Vvecs{k}x}\frac{z}{\gamma_{\Vvecs{k}}}\\
&=&\frac{e}{(2\pi)^3T\sigma_{xx}}\int \frac{dS_{\Vvecs{k}}d\epsilon^*_{\Vvecs{k}}}{z|\Vvec{v}_{\Vvecs{k}}|}\left(-\frac{\partial f}{\partial \epsilon}\right)_{\epsilon=\epsilon^*_{\Vvecs{k}}}\!\!\!\!\!\!\!\!\!\!(\epsilon^*_{\Vvecs{k}})^2\frac{\partial}{\partial k_\perp}\left(\frac{\Vvec{v}_{\Vvecs{k}x}\Vvec{J}_{\Vvecs{k}x}}{|\Vvec{v}_{\Vvecs{k}}|\gamma_{\Vvecs{k}}}\right)_{\epsilon=0}\\
&=&\frac{e\pi^2k_B^2T}{3(2\pi)^3\sigma_{xx}}\int \frac{dS_{\Vvecs{k}}}{z|\Vvec{v}_{\Vvecs{k}}|}\frac{\partial}{\partial k_\perp}\left(\frac{\Vvec{v}_{\Vvecs{k}x}\Vvec{J}_{\Vvecs{k}x}}{|\Vvec{v}_{\Vvecs{k}}|\gamma_{\Vvecs{k}}}\right)_{\epsilon=0}\\
&\propto&z^{-1},
\end{eqnarray}
}
and using $Q_{\Vvecs{k}}\sim
\Vvec{q}_{\Vvecs{k}}=\epsilon\Vvec{v}_{\Vvecs{k}}$ for $T\rightarrow 0$,
$\alpha_{xy}=L^{21}_{xy}(+i\delta)/T^2$ is given by
{\small
\begin{eqnarray}
\alpha_{xy}&\sim&\frac{Be^2}{2(2\pi)^3T}\int \frac{dS_{\Vvecs{k}}d\epsilon^*_{\Vvecs{k}}}{z|\Vvec{v}_{\Vvecs{k}}|}\left(-\frac{\partial f}{\partial \epsilon}\right)_{\epsilon=\epsilon^*_{\Vvecs{k}}}\epsilon^*_{\Vvecs{k}}A''_{\Vvecs{k}}(\epsilon^*_{\Vvecs{k}})\frac{z}{\gamma^2_{\Vvecs{k}}}\\
&=&\frac{Be^2}{2(2\pi)^3T}\int \frac{dS_{\Vvecs{k}}d\epsilon^*_{\Vvecs{k}}}{z|\Vvec{v}_{\Vvecs{k}}|}\left(-\frac{\partial f}{\partial \epsilon}\right)_{\epsilon=\epsilon^*_{\Vvecs{k}}}\!\!\!\!\!\!\!\!\!\!(\epsilon^*_{\Vvecs{k}})^2\frac{\partial}{\partial k_\perp}\left(\frac{A''_{\Vvecs{k}}}{|\Vvec{v}_{\Vvecs{k}}|\gamma^2_{\Vvecs{k}}}\right)_{\epsilon=0}\\
&=&\frac{Be^2\pi^2k_B^2T}{6(2\pi)^3}\int \frac{dS_{\Vvecs{k}}}{z|\Vvec{v}_{\Vvecs{k}}|}\frac{\partial}{\partial k_\perp}\left(\frac{A''_{\Vvecs{k}}}{|\Vvec{v}_{\Vvecs{k}}|\gamma^2_{\Vvecs{k}}}\right)_{\epsilon=0}\\
&\propto&z^{-1},
\end{eqnarray}
}
where $A''_{\Vvecs{k}}(\epsilon)$ is defined by
\begin{equation}
A''_{\Vvecs{k}}(\epsilon)=\gamma_{\Vvecs{k}}(\epsilon)\left\{\Vvec{v}_{\Vvecs{k}}(\epsilon)\times[\Vvec{v}_{\Vvecs{k}}(\epsilon)\times\Vvec{\nabla}]_z\left[\frac{\Vvec{J}_{\Vvecs{k}}(\epsilon)}{\gamma_{\Vvecs{k}}(\epsilon)}\right]\right\}_z
\end{equation}

Thus, we obtain $\nu=\frac{\alpha_{xy}}{B\sigma{xx}}-\frac{S\sigma_{xy}}{B\sigma_{xx}}\propto
z^{-1}$.
We must consider $z$ in detail to calculate $\nu$, because $z$ is much
smaller in heavy fermion systems.
From the experiment of the de Haas-van Alphen\cite{settai}, we can
estimate that the
effective mass $m^*\sim 100m_0$ ($m_0$ is the bare electron mass) and the mass obtained by the band
calculation $m_b\sim 2m_0$ in branches $\beta$.
Then, the mass enhancement factor $z^{-1}$ is given by
$z^{-1}=m^*/m_b\sim 50$.

Although the FLEX can describe various critical
phenomena near the AF QCP, the mass enhancement is not completely
explained with the FLEX in heavy fermion systems.
The reason is that local correlations are not fully taken into account
in the framework of the FLEX, because the vertex corrections in the
self-energy are not included.
According to Ref. \onlinecite{nishikawa}, we separate the
self-energy into the ``local part'' and the ``non-local part''.
In this case, total renormalization factor $z$ is obtained as
$z=z_0z^*$, where $z_0$ is the local renormalization factor which cannot
be included in the FLEX and
 renormalization factor $z^*=(1-(\partial \Sigma/\partial
\omega))^{-1}$ is obtained by the FLEX.
To fit the total renormalization factor to the experimental results ($z^{-1}=50$), we
use $z_0=3/50$, because $z^*\sim 1/3$ in our calculation.

\begin{figure}[htdp]
\begin{center}
\includegraphics[height=60mm]{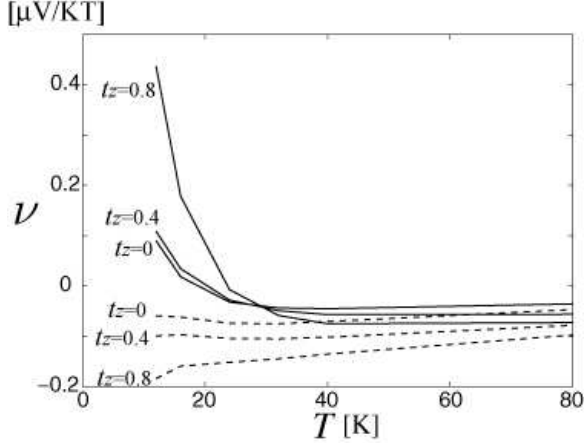}
\caption{The Nernst coefficient $\nu$ with the CVC (solid line) and
 without the CVC (dotted line) for $t_z=0$, $0.4$,
 $0.8$ as a function of $T$ with $\alpha_S=0.995$ at $T=5$K}
\label{ner3}
\end{center}
\end{figure}

We show the obtained temperature dependence of the Nernst coefficient
$\nu$ in Fig. \ref{ner3}, where solid and dotted lines correspond to
$\nu$ with and without the CVC, respectively, for $t_z=0$,
$t_z=0.4$ and $t_z=0.8$.
In this figure, we chose the parameters for heavy fermion CeCoIn$_5$,
i.e., $a_a=a_b=4.6\times10^{-10}$m, and then, $\frac{k_Ba_aa_b}{\hbar}\sim 28\rm{nV/KT}$ has been
multiplied as a unit of calculated value, and the local mass enhancement
factor $z_0^{-1}$ has been also multiplied. 

\begin{figure}[htdp]
\begin{center}
\includegraphics[height=60mm]{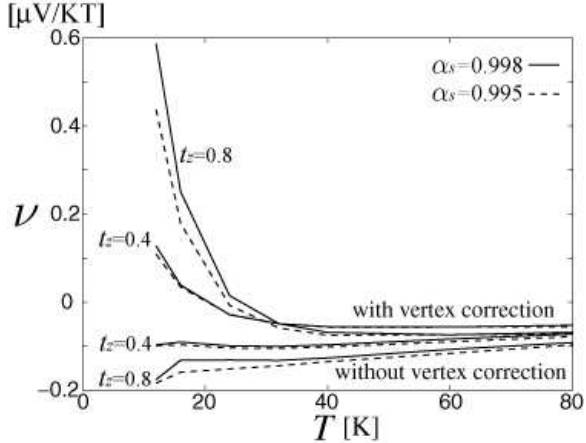}
\caption{The Nernst coefficient $\nu$ with the CVC (upside) and
 without the CVC (downside) for $t_z=0.4$,
 $0.8$ as a function of $T$ with  $\alpha_S=0.995$ (solid line) and $0.998$ (dotted line) at $T=5$K.}
\label{ner-2}
\end{center}
\end{figure}

We see that $\nu$ without the CVC is almost constant, and $\nu$ with the
CVC shows enormous increase
at low temperatures, especially in strong three-dimensional case
($t_z=0.8$) where the Fermi surface is similar to that of CeCoIn$_5$. Then, the temperature dependence of $\nu$
resembles to that of $R_H$. This temperature dependence of $|\nu|$ is consistent with
the giant Nernst effect in CeCoIn$_5$ ($|\nu|\sim 1\mu$V/KT for $T=5$K).\cite{bel}
Here, we discuss the reason why the magnitude of $\nu$ becomes large.
$A_{\Vvecs{k}}$ and $A'_{\Vvecs{k}}$ given in eqs. (\ref{A}) and (\ref{A'}) are rewritten as
\begin{eqnarray}
A_{\Vvecs{k}}(\epsilon)&=&|\Vvec{v}_{\Vvecs{k}}(\epsilon)|_\perp\left(\Vvec{J}_{\Vvecs{k}}(\epsilon)\times \frac{\partial}{\partial k_\parallel}\Vvec{J}_{\Vvecs{k}}(\epsilon)\right)_z\\
&=&|\Vvec{v}_{\Vvecs{k}}(\epsilon)|_\perp|\Vvec{J}_{\Vvecs{k}}(\epsilon)|_\perp^2\frac{\partial\theta_{\Vvecs{J}}(\Vvec{k})}{\partial k_\parallel},\\
A'_{\Vvecs{k}}(\epsilon)&=&\gamma_{\Vvecs{k}}(\epsilon)|\Vvec{v}_{\Vvecs{k}}(\epsilon)|_\perp\left(\Vvec{Q}_{\Vvecs{k}}(\epsilon)\times \frac{\partial}{\partial k_\parallel}\frac{\Vvec{J}_{\Vvecs{k}}(\epsilon)}{\gamma_{\Vvecs{k}}(\epsilon)}\right)_z\\
&=&|\Vvec{v}_{\Vvecs{k}}(\epsilon)|_\perp\left[(Q_{\Vvecs{k}x}J_{\Vvecs{k}x}+Q_{\Vvecs{k}y}J_{\Vvecs{k}y})\frac{\partial\theta_{\Vvecs{J}}(\Vvec{k})}{\partial k_\parallel}\right.\nonumber\\
&&\left.+(\Vvec{Q}_{\Vvecs{k}}\times \Vvec{J}_{\Vvecs{k}})_z\frac{\partial}{\partial k_\parallel}\log\left(\frac{|\Vvec{J}_{\Vvecs{k}}|_\perp}{\gamma_{\Vvecs{k}}}\right)\right].
\end{eqnarray}
Here, $\Vvec{Q}_{\Vvecs{k}}$ is the total heat current with the CVC.
We stress that $\Vvec{Q}_{\Vvecs{k}}\sim \epsilon\Vvec{v}_{\Vvecs{k}}$
and $\Vvec{Q}_{\Vvecs{k}}$ is not parallel to $\Vvec{J}_{\Vvecs{k}}$
when the AF fluctuations are strong.\cite{kontani-nu}
In this case, $\nu$ is strongly enhanced due to the second term of $A'_{\Vvecs{k}}$ in
high-$T_c$ cuprates.\cite{kontani-yamada}
We expect that the same mechanism will give the enhancement of $\nu$ in
the present study for three-dimensional case.
In NCCO, $\nu$ is enhanced by the CVC due to
the AF fluctuation below $300$K,
whereas the increment of $\nu$ for LSCO is brought by the CVC
due to both AF and superconducting fluctuations
below $T^*\sim 150$K.
 \cite{kontani-nu}
Because of the relation $\nu\propto\gamma^{-1}$, $\nu$ is proportional to
$\rho^{-1}$ for a fixed $\alpha_S$. This fact would contribute to the
enhancement of $\nu$ in 3D case, as shown in Figs. \ref{ner3}, \ref{ner-2}.
We show the temperature dependence of $\nu$ for both
$\alpha_S=0.995$ and $0.998$ in Fig. \ref{ner-2}.
$\nu$ with the CVC increases as $\alpha_S$ approaches unity, which resembles to the behavior of
$R_H$.
Consequently, the magnitude of $\nu$ increases almost divergently in the vicinity of the AF QCP.

%As for the absolute value of $\nu$, comparing to the experimental result for heavy fermion
%CeCoIn$_5$ $(|\nu|\sim 1\mu V/KT)$, our results remain quite small.
%We can improve this discrepancy considering the effect of the
%renormalization factor $Z$. From the relation $\nu\propto
%1/(Z\gamma)$, we expect that $\nu$ for lower temperature ($T\sim 0.02$)
%reaches $1\mu V/KT$ if $Z$ in CeCoIn$_5$ is about 10 times smaller than
%that in high-$T_c$ cuprate. It is reasonable to estimate the $Z$ for
%heavy fermion is about 10 times smaller than
%that in high-$T_c$ cuprate, because mass for heavy fermion is about 10
%times larger than that for high-$T_c$ cuprate. 

\section{Conclusion}
We have calculated microscopically the resistivity $\rho$, the Hall coefficient
$R_H$ and Nernst
coefficient $\nu$ for three-dimensional Hubbard model close to the AF QCP
based on the Fermi-liquid theory. This is a first microscopic
calculation for the Hall coefficient and the Nernst coefficient with the
current vertex corrections (CVC) in the
three-dimensional system. 
In two-dimensional systems, it is established that the CVC plays an important role when the AF fluctuations are strong.
On the other hand, the CVC vanishes completely in infinite dimension
$d=\infty$. Thus, it is a very important theoretical issue to clarify
whether the CVC is significant or not in three-dimensional systems.
We find that the CVC influences crucially on various transport phenomena
in both two and three-dimensional systems close to the AF QCP.

We have shown that the magnitude of $R_H$ and $\nu$ is strongly enhanced
with the decrease of temperature due to the CVC. 
These strong temperature dependences in the Hall coefficient and the Nernst coefficient 
come from the difference between the direction of the total current
$(\Vvec{J}_{\Vvecs{k}})$ and that of $\Vvec{v}_{\Vvecs{k}}$
around the cold spots. The difference of directions increases as the
temperature decreases near the
QCP, which can be expressed in terms of the effective curvature $\partial
\theta_{\Vvecs{J}}/\partial k_\parallel$ of
Fermi surface obtained by the direction of $\Vvec{J}_{\Vvecs{k}}$.
 The obtained values of $R_H$ at the lowest temperature ($T=0.02$) is more than $5$ times larger
than those at high temperatures for three-dimensional system ($t_z=0.8$). 
This result is qualitatively consistent with experimental results in various three-dimensional heavy fermion
systems close to the AF QCP, such as CeCoIn$_5$ and
CeCu$_{6-x}$Au$_x$.
This strong enhancement cannot be
explained by the relaxation time approximation (RTA).

In the present paper, we also studied the momentum dependence of relaxation time
in three-dimensional systems due to strong AF spin
fluctuations. In two-dimensional systems, it is known that hot spots and
cold spots take line structures along $z$ axis, as the systems approach
to the AF QCP.
In three-dimensional systems, we find that hot spots become point-like
(``hot points'')
while the cold spots remain to take line structures (``cold lines'').
The emergence of hot points and cold lines is expected to be general in
three-dimensional systems close to the AF QCP.
Transport phenomena are mainly determined by the cold spots.
The area of cold spots in the phase space plays an important role.
We find that the CVC around cold spots produces strong enhancement
of $R_H$ and $\nu$.
We emphasize that the strong enhancement of $R_H$ and $\nu$ comes from the effective curvature of the Fermi surface, $\partial
\theta_{\Vvecs{j}}/\partial k_\parallel$, enhanced by the CVC on cold
lines, as shown in Fig. \ref{ang-tz0.8}.
Note that obtained results for $R_H$ and $\nu$ without the CVC, which corresponds to
the RTA, are almost temperature independent.

In future, we will perform a quantitative study for the transport phenomena in CeCoIn$_5$ and
CeRhIn$_5$, using a realistic band structure predicted by band
calculations.\cite{maehira,harima} 
In the present paper, signs of $R_H$ and $\nu$ are
opposite to actual experimental results.
We expect that this discrepancy can be resolved by taking into account
a proper band structure.

\section{ACKNOWLEDGMENTS}
This work was supported by a Grant-in-Aid for 21st Century COE
``Frontiers of Computational Science''.
Numerical calculations were performed at the supercomputer center, ISSP.
The authors are grateful to K. Yamada, J. Inoue, N. Nagaosa, Y. Matsuda,
Y. Suzumura, D. Hirashima, Y. Nakajima and K. Tanaka for useful comments and discussions. 

\bibliography{paper}

\end{document}